\documentclass[jkps,preprint,fleqn,showpacs,showkeys]{revtex4}
\usepackage{graphicx}
\usepackage{amssymb}
\usepackage{amsmath}
\usepackage{bm}
\begin{document}
\setcounter{page}{0}
\title[]{Dirac Phenomenological Analyses of 1.047 GeV Proton Inelastic Scatterings from $^{62}$Ni and $^{64}$Ni}	
\author{ Sugie \surname{Shim}}
\email{shim@kongju.ac.kr}
\thanks{Fax: +82-41-850-8489}
\affiliation{Department of Physics, Kongju National University, Gongju 32588}

\date[]{Received 2018}

\begin{abstract}
Unpolarized 1.047 GeV proton inelastic scatterings from Ni isotopes, $^{62}$Ni and $^{64}$Ni are analyzed phenomenologically employing an optical potential model and the first order collective model in the relativistic Dirac coupled channel formalism. The Dirac equations are reduced to the Schr\"{o}dinger-like second-order differential equations and the effective central and spin-orbit optical potentials are analyzed by considering mass number dependence. The multistep excitation via $2^+$ state is found to be important for the $4^+$ state excitation in the ground state rotational band at the proton inelastic scatterings from the Ni isotopes. The calculated deformation parameters for the 2$^+$, the 4$^+$ states of the ground state rotational band and the first 3$^-$ state are found to agree pretty well with those obtained in the nonrelativistic calculations.
\end{abstract}



\maketitle

\section{INTRODUCTION}

Relativistic Dirac approaches based on the Dirac equation have been very successful for describing the intermediate energy proton scatterings from the nuclei, achieving better agreement with the experimental data than the nonrelativistic approaches based on the Schr\"{o}dinger equation \cite{1,2,3,4,5,6,7,8,9,10}. However, it is still necessary to analyze more nuclear scattering data using the Dirac approach in order to complete the systematic Dirac analyses and eventually to provide a reliable basis for replacing the nonrelativistic Schr\"{o}dinger approach with the relativistic Dirac approach for the analyses of the nuclear scatterings.

In this work we performed a relativistic Dirac coupled channel analysis for the inelastic proton scatterings from Ni isotopes, $^{62}$Ni and $^{64}$Ni, by using an optical potential model \cite{1} and the first order collective model. This work is a follow-up of our previous publication for the Dirac phenomenological analyses of the inelastic proton scatterings from the other Ni isotopes, $^{58}$Ni and $^{60}$Ni \cite{11} . Ni isotopes are of interest because they are known to have a doubly closed shell ($N=Z=28$) surrounded by only a few off-shell neutrons \cite{12}.
The Dirac optical potential and the deformation parameters are searched to fit the experimental data using a computer program called ECIS \cite{13}, where a Numerov method is employed to solve the complicated Dirac coupled channel equations.  The Dirac equations are reduced to the Schr\"{o}dinger-like second-order differential equations and the effective central and spin-orbit optical potentials are analyzed by considering the mass number dependence.

\section{Theory and Results}

Dirac phenomenological analyses are performed for the 1.047 GeV unpolarized proton inelastic scatterings from Ni isotopes, $^{62}$Ni and $^{64}$Ni, by employing an optical potential model and a first-order collective model. Ni isotopes are of interest because they have the closed proton shell, Z=28. They are known to have a closed 1 $f_{7/2}$ proton shell with a few off-shell neutrons outside the closed neutron 1 $f_{7/2}$ shell \cite{12}.
$^{62}$Ni and $^{64}$Ni are spin-0 nuclei and most of the theoretical procedures for the Dirac phenomenological calculation for the proton scatterings from spin-0 nuclei are given in our previous publications \cite{3,4,8,9,10, 11,14,15}. Hence, they will be omitted in this paper.
 The Dirac equation may be rewritten as two coupled equations for the upper ($\Psi_u$) and lower ($\Psi_l$) components of the Dirac wave function, $\Psi (r)$, and we let
\begin{equation}
 \Psi_u (r) = K(r)\psi(r) , \hspace{.5in} K(r)=A^{1/2} \exp [\int iU_V^r (r) dr ]
\label{e3}
\end{equation}
where $K(r)\rightarrow 1$ as $r \rightarrow \infty $, $A=(m+U_S +E-U_V^0 )/(m+E) $. Here, $U_S$ is the scalar potential, $U_V^r$ and $U_V^0$ are the space-like and the time-like vector potentials, respectively.
Under this wave function transformation, we can have the Schr\"{o}dinger-like second-order differential equation for $\psi(r)$ as follows and can compare with the conventional nonrelativistic Schr\"{o}dinger equation.
  \begin{equation}
[p^2 + 2E(U_{cent} +U_{SO} {\bf \sigma} \cdot {\bf L} ) ]\psi
(r)=[(E-V_c )^2- m^2 - \frac{2U_{AM} }{r} -\frac{\partial U_{AM} }{\partial r} -U_{AM}^2] \psi (r).
\label{e3}
\end{equation}
Here, the Schr\"{o}dinger
 equivalent or effective central potential which contains the Darwin potentials, and effective spin-orbit potentials are defined as follows.
\begin{eqnarray}
U_{cent} & = & \frac{1}{2E} [2EU_V^0 +2mU_S -U_V^{02}+U_S^2-2V_c U_V^0 \nonumber \\
& & +U_T^2 +2U_T
U_{AM}-\frac{U_T+U_{AM}}{A}(\frac{\partial A}{\partial r}) \nonumber \\
& & +\frac{2U_T}{r}+2EU_{Darwin}] \nonumber \\
U_{Darwin} & = & \frac{1}{2E} [-\frac{1}{2r^2 A} \frac{\partial }{\partial r}(r^2 \frac{\partial A}{\partial r})+\frac{3}{4A^2}(\frac{\partial A}{\partial r})^2] \nonumber \\
U_{SO} & = & \frac{1}{2E}[\frac{1}{r A}(\frac{\partial A}{\partial r})+\frac{2}{r}(U_T+U_{AM})]
\end{eqnarray}
Here, $U_{AM}(r)=\frac{k}{2m} \frac{\partial}{\partial r} V_c (r)$ and $k$ is the abnormal magnetic moment ($k=1.79$ for proton, $k=-1.91$ for neutron).
 Hence in the Dirac approach, it is shown that the spin-orbit potential appears naturally when we reduce the Dirac equation to a Schr\"{o}dinger-like second-order differential equation, while in the nonrelativistic Schr\"{o}dinger approach, we have to insert the spin-orbit potential by hand.

The Dirac equations are numerically solved to get the parameters fitting best to the experimental data by employing the minimum $\chi^2$ method.
In order to obtain the optimizing optical potential parameter set we minimize the chi-square for given scattering observables
by varying the adjustable parameters in the coupled differential equations and iterations. When the number of experimental data is $n$
for the given angular distribution of scattering observables, the chi-square $\chi^2 $ is defined as
\begin{equation}
\chi^2 = \sum_{i=1}^n {|x_{th} (\theta_i )-x_{exp} (\theta_i )|^2 \over (\Delta x_{exp} (\theta_i ) )^2 } ,
\end{equation}

where $x_{th} $ denotes the theoretical value, $x_{exp} $ denotes the experimental value and $\Delta x_{exp} $ denotes the experimental error of the scattering observable which is the scattering differential cross section in this work.

The experimental data are obtained from Ref. 16 for the 1.047 GeV unpolarized proton inelastic scatterings from $^{62}$Ni and $^{64}$Ni.
The first $2^+$ and $4^+$ states are assumed to be members of the ground state rotational band (GSRB) ($J^{\pi}=0^+$) and also assumed to be collective rotational states.
As a first step, the 12 parameters of the direct scalar and vector potentials in Woods-Saxon shapes are searched to reproduce the elastic scattering experimental data.
The calculated results are shown as dotted lines in Figs. 1 and 2 for the elastic scatterings from $^{62}$Ni and $^{64}$Ni, respectively. It is seen that the results of the Dirac phenomenological calculations can reproduce the elastic experimental data quite well, showing better agreement with the data compared to the results obtained in the nonrelativistic calculations \cite{16}. In the figures, `cpd' means `coupled'.

\begin{figure}
\includegraphics[width=10.0cm]{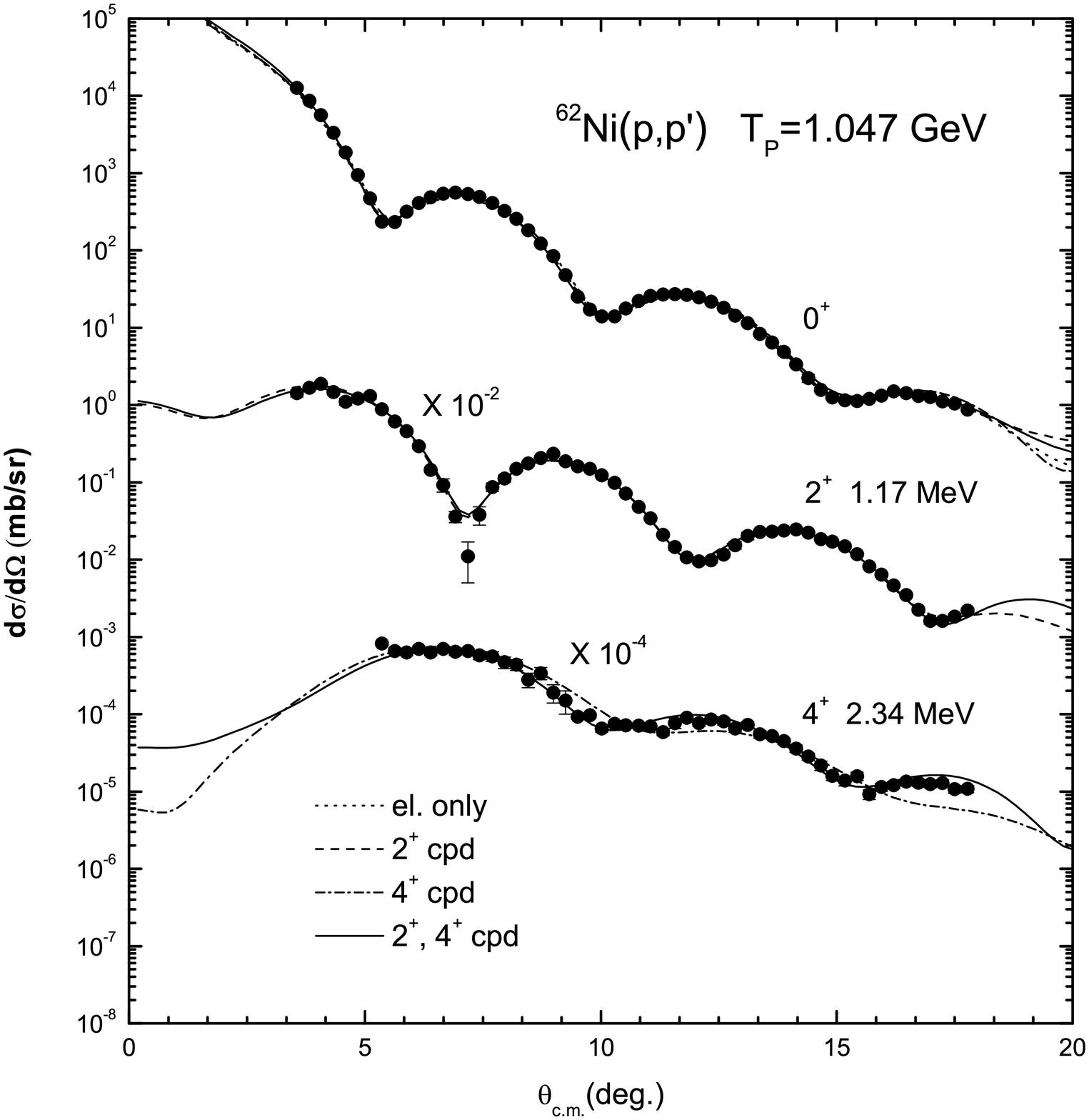}
\caption[0]{Differential cross section of the low-lying excited states of the GSRB for 1.047 GeV p +  $^{62}$Ni scattering. The dotted, dashed, dash-dot and solid lines represent the results of Dirac calculation where only the elastic scattering is considered, where the ground state and the $2^+$ state are coupled, where the ground state and the $4^+$ state are coupled, and where the ground state, the $2^+$ state and the $4^+$ state are coupled, respectively.}
\label{fig1}
\end{figure}

\begin{figure}
\includegraphics[width=10.0cm]{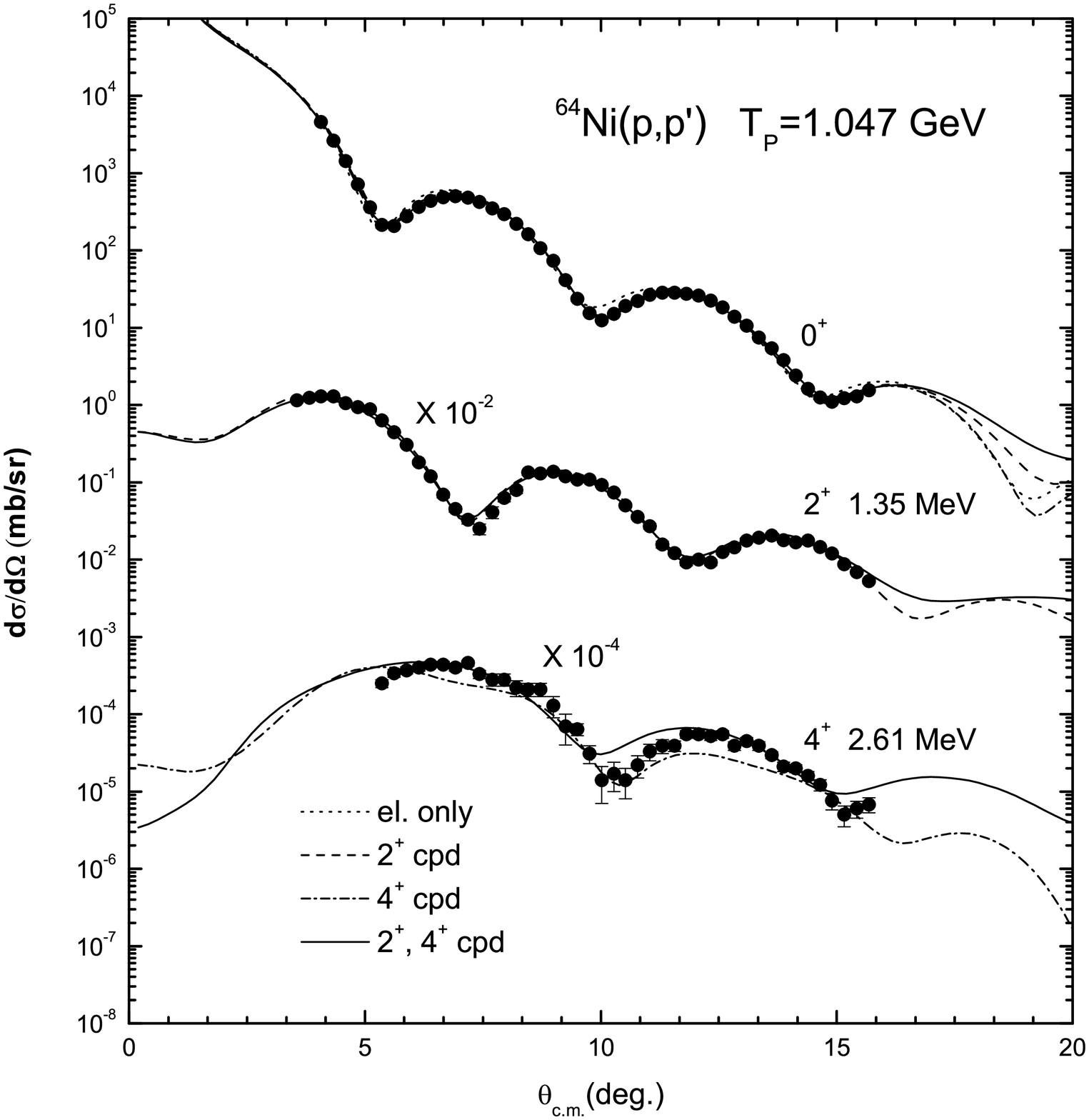}
\caption[0]{Differential cross section of the low-lying excited states of the GSRB for 1.047 GeV p +  $^{64}$Ni scattering. The dotted, dashed, dash-dot and solid lines represent the results of Dirac calculation where only elastic scattering is considered, where the ground state and the $2^+$ state are coupled, where the ground state and the $4^+$ state are coupled, and where the ground state, the $2^+$ state and the $4^+$ state are coupled, respectively.}
\label{fig1}
\end{figure}

\begin{table}
\caption{Calculated optical potential parameters of a Woods-Saxon shape for 1.047 GeV proton elastic scatterings from $^{62}$Ni and $^{64}$Ni.}
\begin{ruledtabular}
\begin{tabular}{ccccccccc}

   Potential & Nucleus     &   Strength (MeV)   & Radius (fm)  & Diffusiveness (fm)  ~ \\
   \hline
 Scalar & $^{62}$Ni & -356.5     & 3.284 & 0.6815        ~ \\
 real   & $^{64}$Ni & -38.56     &  7.204   & 0.8492           ~ \\ \hline
 Scalar  &$^{62}$Ni & 941.2     & 3.324 & 0.8398        ~ \\
 imaginary    & $^{64}$Ni & 1.822    &  5.109   &  0.5247          ~ \\ \hline
 Vector  & $^{62}$Ni &402.5    & 3.019 & 0.6310         ~ \\
 real    & $^{64}$Ni &  22.58   &  6.817   & 1.0310          ~ \\ \hline
 Vector  & $^{62}$Ni &-407.7     & 3.583 & 0.6141       ~ \\
 imaginary  &$^{64}$Ni &-90.48      &  4.121   &  0.6189           ~ \\
\end{tabular}
\end{ruledtabular}
\label{table1}
\end{table}

\begin{table}
\caption{Calculated optical potential parameters of a Woods-Saxon shape for 1.047 GeV proton inelastic scatterings from $^{62}$Ni and $^{64}$Ni, for the cases where all three states, $0^+$, $2^+$ and $4^+$ states, are coupled.}
\begin{ruledtabular}
\begin{tabular}{ccccccccc}

   Potential & Nucleus     &   Strength (MeV)   & Radius (fm)  & Diffusiveness (fm)  ~ \\
   \hline
 Scalar & $^{62}$Ni & -34.24     & 5.088 & 0.4274        ~ \\
 real   & $^{64}$Ni & -127.9     &  6.653   & 0.9248           ~ \\ \hline
 Scalar  &$^{62}$Ni & 948.2     & 3.383 & 0.4829        ~ \\
 imaginary    & $^{64}$Ni & 72.50    &  5.022   &  0.2833          ~ \\ \hline
 Vector  & $^{62}$Ni &256.3    & 3.067 & 0.6277         ~ \\
 real    & $^{64}$Ni &  63.79   &  6.562   & 1.0013          ~ \\ \hline
 Vector  & $^{62}$Ni &-546.8     & 3.391 & 0.5754       ~ \\
 imaginary  &$^{64}$Ni &-92.97      &  4.670   &  0.4454           ~ \\
\end{tabular}
\end{ruledtabular}
\label{table1}
\end{table}

The calculated optical potential parameters of the Woods-Saxon shape for the 1.047 GeV proton scatterings from $^{62}$Ni and $^{64}$Ni are shown in Tables I and II for the elastic scattering and for the inelastic scattering where all three states of the GSRB, $0^+$, $2^+$ and $4^+$ states are coupled, respectively.
 Showing the similar pattern as in spherically symmetric nuclei \cite{3}, the real scalar potentials and the imaginary vector potentials are found to be large and negative, and that the imaginary scalar potentials and the real vector potentials to be positive and large, except the imaginary scalar potential for the elastic scattering from $^{64}$Ni which is found to be rather small.
It is observed that the strength parameters of all four potentials mostly decrease as the mass number is increased from 62 to 64, for both elastic and inelastic scatterings, except at the real scalar potentials when the inelastic scattering is considered. The radius parameters of the potentials increase as the mass number is increased from 62 to 64, as expected.
As a first step for inelastic scattering calculations, only the ground state and one excited
state, the $2^+$ state or the $4^+$ state, are included at once in the calculations. Next, the
ground state, the $2^+$ state, and the $4^+$ state are  included in the inelastic scattering calculations to investigate
the effect of the channel coupling between the excited states of the GSRB, which is known to be strong as shown in our previous publications for the proton scatterings from axially symmetric deformed nuclei \cite{9, 11}.
The Dirac coupled channel equations are solved phenomenologically to obtain the best fitting optical potential and deformation parameters to the experimental data by using the minimum $\chi^2$ method.
The real and the imaginary $\beta_\lambda$ are
set to be  equal for a given potential type, so that $\beta_S$ and
$\beta_V$ are
determined for each excited state.
In Figs. 1 and 2, the calculated results for the the $2^+$ state and the $4^+$ state are also shown. For the $2^+$ state, the agreement with the experimental data didn't change noticably by adding the coupling with the $4^+$ state in the calculation.
We observe that the $\chi^2$ for the $2^+$ state is
reduced slightly
when the coupling with the $4^+$ state is added in the calculation.
However, the agreement with the experimental data for the $4^+$ state improved significantly by adding the coupling with the $2^+$ state in the calculation, indicating multistep excitation via $2^+$ state is important for the $4^+$ state excitation in the GSRB at the proton scatterings from both nuclei, $^{62}$Ni and $^{64}$Ni, which is the same feature found at the scatterings from $^{58}$Ni and $^{60}$Ni \cite{11}.
$\chi^2/n$ for the $4^+$ state is
reduced to about 1/3, from 10.02 to 3.12 for the case of $^{62}$Ni, and from 10.03 to 3.09 for the case of $^{64}$Ni,
when the coupling with the $2^+$ state is added in the calculations. However, it is observed that the theoretical values are shifted a little from the data at the first and the second minima of the $4^+$ state data at the scattering from $^{64}$Ni. It can be due to the coupling effect with the $6^+$ state which possibly belong to the GSRB or the higher level excitations near the $4^+$ excitation energy level, which are not included in this calculation.

\begin{figure}
\includegraphics[width=10.0cm]{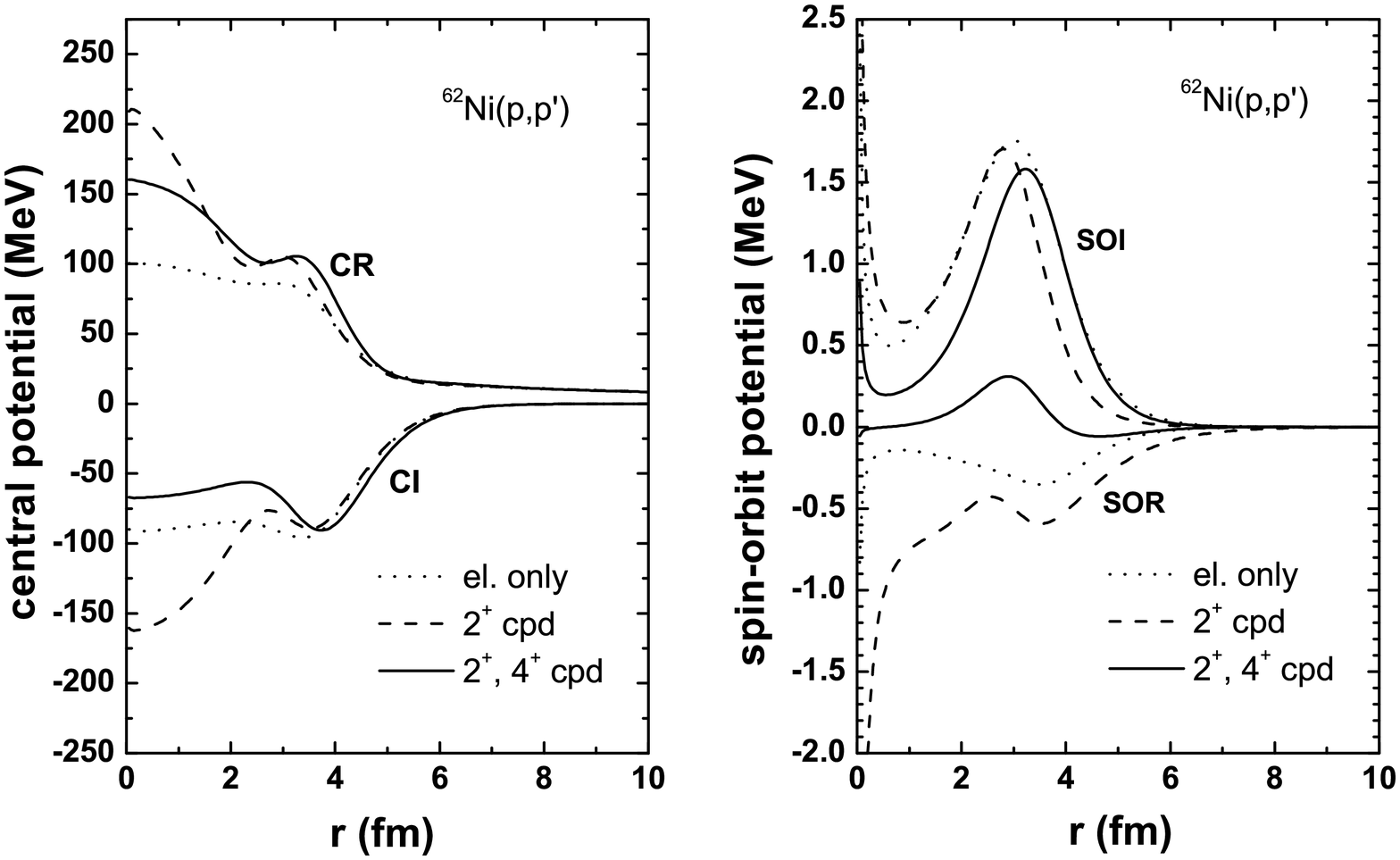}
\caption[0]{The effective central and spin-orbit potentials for the proton scattering from $^{62}$Ni. CR and CI denote  central real and imaginary optical potentials, and SOR and SOI denote  spin-orbit real and imaginary optical
potentials, respectively.}
\label{fig2}
\end{figure}

\begin{figure}
\includegraphics[width=10.0cm]{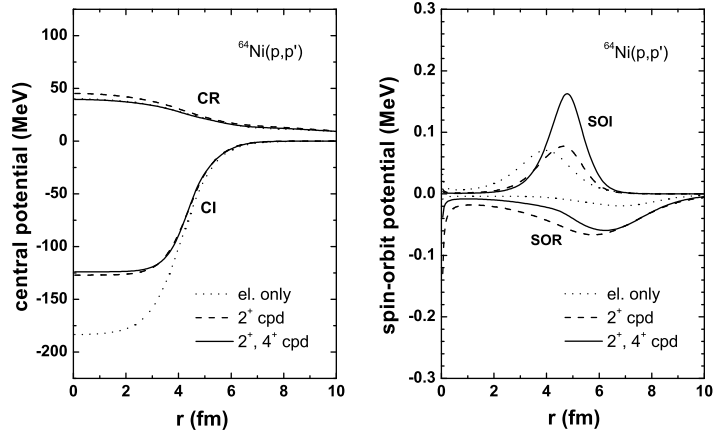}
\caption[0]{The effective central and spin-orbit potentials for the proton scattering from  $^{64}$Ni. CR and CI denote  central real and imaginary optical potentials, and SOR and SOI denote  spin-orbit real and imaginary optical
potentials, respectively.}
\label{fig2}
\end{figure}

In Figs. 3 and 4, the effective central and spin-orbit potentials for the proton scatterings from $^{62}$Ni and $^{64}$Ni are shown. The dotted, dashed, and solid lines represent the results of the Dirac phenomenological calculations where only elastic scattering is considered, where the ground state and the $2^+$ state are coupled, and where the ground state, the $2^+$ state and the $4^+$ state are coupled, respectively. Surface-peaked phenomena are observed at the effective central potentials for the scattering from $^{62}$Ni as shown at the scatterings from the other axially deformed nuclei such as $^{20}$Ne and $^{24}$Mg \cite{8, 14}, whereas the surface-peaked phenomena are not observed at the effective central potentials for the scattering from  $^{64}$Ni. The effective central potentials are observed to have about the same values near the surface, at near 4 fermi, for all three cases, at both nuclei, while the spin-orbit potentials are found to have a little variations near the surface.
The surface-peaked phenomena are shown at the all effective spin-orbit potentials, indicating that the spin-orbit interaction may be considered as a surface-peaked interaction.
Somehow, it is observed that the peak position of the imaginary spin-orbit potential is found at near 3 fermi for the scattering from $^{62}$Ni, whereas the peak position of real spin-orbit potential is found at near 6 fermi for the scattering from $^{64}$Ni.

\begin{figure}
\includegraphics[width=10.0cm]{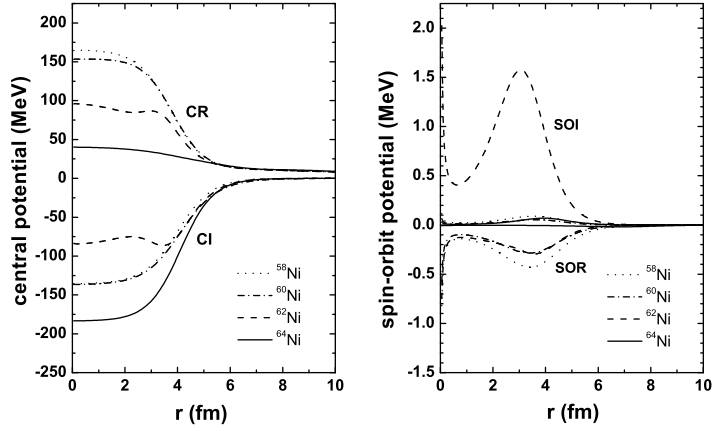}
\caption[0]{The effective central and spin-orbit optical potentials for the proton elastic scattering from Ni isotopes.}
\label{fig2}
\end{figure}

\begin{figure}
\includegraphics[width=10.0cm]{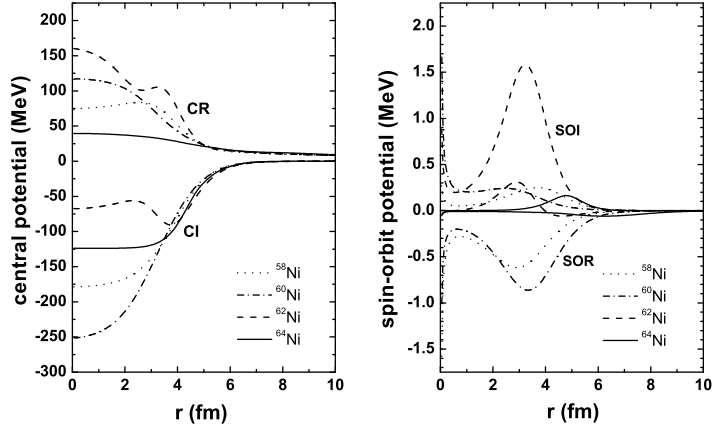}
\caption[0]{The effective central and spin-orbit optical potentials for the proton inelastic scattering from  Ni isotopes, for the case where the ground state, the $2^+$ state and the $4^+$ state are coupled.}
\label{fig2}
\end{figure}

In Fig. 5, the effective central and spin-orbit potentials for the proton elastic scattering from Ni isotopes, $^{58}$Ni, $^{60}$Ni \cite{11}, $^{62}$Ni and $^{64}$Ni are compared with each other. In Fig. 6 the effective potentials for the proton inelastic scattering from the Ni isotopes are shown for the case where the ground state, the $2^+$ state and the $4^+$ state are coupled. The dotted, dash-dot, dashed, and solid lines represent the results of the Dirac phenomenological calculations for the proton scatterings from  $^{58}$Ni, $^{60}$Ni, $^{62}$Ni and $^{64}$Ni, respectively.
It is shown that the peak position of the effective real spin-orbit potential is moved to the direction of large $r$ as the mass number is increased, but the tendency is not shown clearly at the effective imaginary spin-orbit potentials, for both cases, as shown in Figs. 5 and 6. The strength parameters of the real effective central and spin-orbit potentials decrease as the mass number is increased for the elastic scattering, but the tendency is not shown for the inelastic scatterings. The real and the imaginary parts of the effective central potentials and the real parts of the spin-orbit potentials for the scattering from $^{62}$Ni are observed to have abnormal wiggling shapes at near 3 fermi, indicating some inner structure of the nucleus.

\begin{figure}
\includegraphics[width=10.0cm]{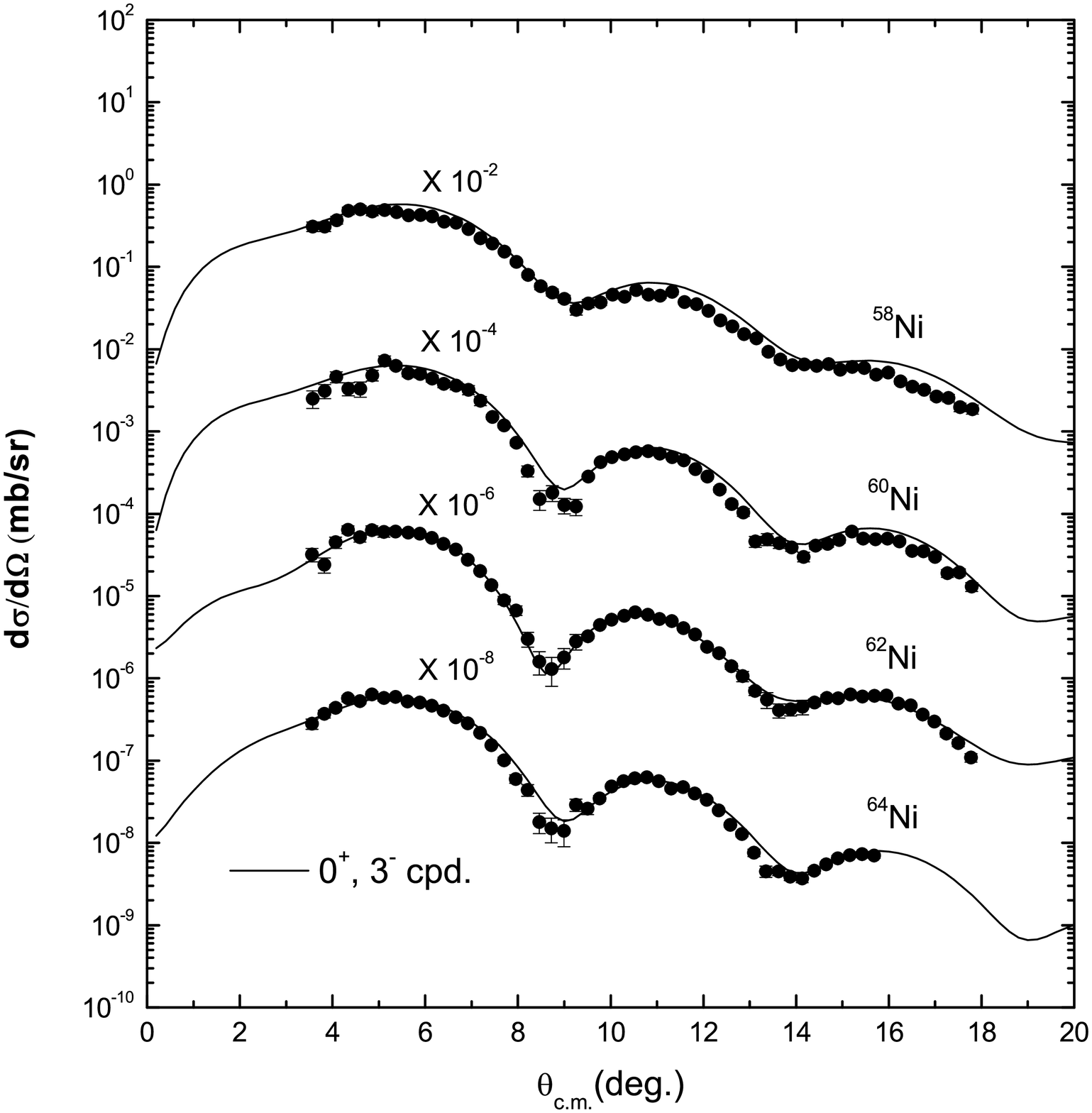}
\caption[0]{Differential cross section of the first 3$^-$ states of the 1.047 GeV proton scatterings from Ni isotopes. The solid lines represent the results of Dirac calculation where the ground state and the $3^-$ state are coupled.}
\label{fig1}
\end{figure}

\begin{figure}
\includegraphics[width=10.0cm]{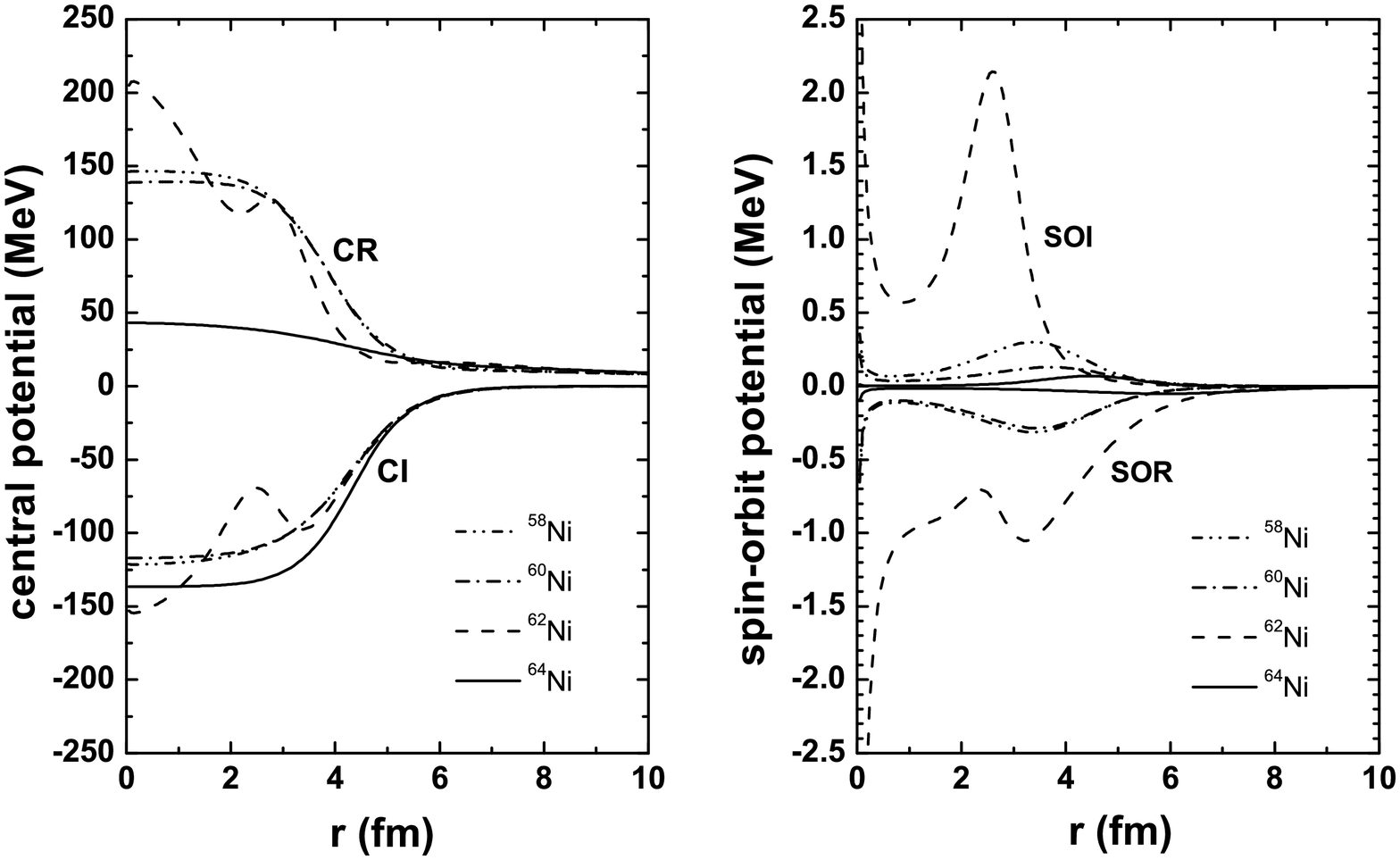}
\caption[0]{The effective central and spin-orbit optical potentials for the proton inelastic scattering from  Ni isotopes, for the case where the ground state and the $3^-$ state are coupled.}
\label{fig2}
\end{figure}

\begin{table}
\caption{The calculated deformation parameters for the $2^+ $ states and the $4^+$ states for 1.047 GeV proton scatterings from Ni isotopes are shown for the case where the ground state, the $2^+$ state and the $4^+$ state are coupled.
The calculated deformation parameters for the $3^- $ states for 1.047 GeV proton scatterings from Ni isotopes are also shown for the case where the ground state and the $3^-$ state are coupled.}
\begin{ruledtabular}
\begin{tabular}{c|ccccccc}
   &Target       &  Energy   &   &   &   ~ \\
   & nuclei      &  (MeV)  & $\beta_S $  & $\beta_V $  & $\beta_{NR} $  ~ \\
   \hline \hline
  & $ ^{58}Ni $ & 1.45  &   0.24   &  0.21   &  $0.233^{17}, 0.187^{18}, 0.207^{19} $       ~ \\ \cline{2-6}
 $2^+ $ state & $^{60}Ni $  & 1.33    & 0.25 & 0.24    &  $0.211^{18}, 0.232^{19}, 0.255^{21} $     ~ \\ \cline{2-6}
 & $^{62}Ni $  & 1.17    & 0.209 & 0.233    &   $0.193^{18}, 0.26^{12} $   ~ \\ \cline{2-6}
 & $^{64}Ni $    &  1.35    &  0.188   &   0.199  & $0.192^{18}, 0.22^{12}, 0.206^{21} $         ~ \\ \hline \hline
   & $ ^{58}Ni $ & 2.46  &   0.07   &  0.08   &  $0.093^{17}, 0.10^{20} $       ~ \\ \cline{2-6}
 $4^+ $ state & $^{60}Ni $  & 2.50    & 0.11 & 0.10    & $ 0.127^{21}$    ~ \\ \cline{2-6}
 & $^{62}Ni $  & 2.34    & 0.037 & 0.054    & $ 0.11^{12}  $    ~ \\ \cline{2-6}
 & $^{64}Ni $    &  2.61    &  0.046   &   0.051  &  $ 0.09^{12} $       ~ \\  \hline \hline
   & $ ^{58}Ni $ & 4.47  &   0.180   &  0.160   &  $0.173^{19} $       ~ \\ \cline{2-6}
 $3^- $ state & $^{60}Ni $  & 4.04    & 0.192 & 0.181    & $ 0.186^{19}, 0.209^{21}$    ~ \\ \cline{2-6}
 & $^{62}Ni $  & 3.76    & 0.206 & 0.194    & $ 0.23^{12}  $    ~ \\ \cline{2-6}
 & $^{64}Ni $    &  3.55    &  0.180   &   0.191  &  $0.23^{12}, 0.203^{21} $       ~ \\
\end{tabular}
\end{ruledtabular}
\label{table2}
\end{table}

In Table III, we show the deformation parameters for the $2^+$ states and the $4^+$ states of Ni isotopes. It also contains the results of our previous calculations for the proton scatterings from $^{58}$Ni and $^{60}$Ni \cite{11}. It is shown that the deformation parameters for the $2^+$ state at $^{64}$Ni are smaller than those at $^{62}$Ni, as expected from the fact that the excitation energy of the state is larger at the scattering from  $^{64}$Ni. We can say that the $2^+$ state is less strongly coupled to the ground state at the scattering from $^{64}$Ni than at the scattering from $^{62}$Ni. However, the deformation parameter $\beta_S $ for the $4^+$ state excitation at the scattering from $^{62}$Ni is found to be smaller than that of $^{64}$Ni, even when the excitation energy is smaller at $^{62}$Ni.  It is found that $\beta_V$ is lager than $\beta_S$ for the scatterings from $^{62}$Ni and $^{64}$Ni nuclei, while the tendency is true only for the 4$^+$ state excitation at the scattering from $^{58}$Ni when the scatterings from $^{58}$Ni and $^{60}$Ni are considered.  We also performed the Dirac phenomenological calculation for the inelastic scatterings from the Ni isotopes considering the first 3$^-$ excitation by using the first order vibrational collective model. The calculated results for the differential cross-section is shown in Fig. 7 and the effective potentials for the 3$^-$ state coupled case are shown in Fig. 8. It is clearly shown that the results of the Dirac phenomenological calculations give better agreement with the experimental data compared to those obtained in the nonrelativistic calculations \cite{16}. The deformation parameters for the $3^- $ states for 1.047 GeV proton scatterings from Ni isotopes are also shown for the case where the ground state and the $3^-$ state are coupled, in Table III. It is found that the deformation parameters for the 2$^+$, the 4$^+$ states of the GSRB and the first 3$^-$ state agree pretty well with those obtained in the nonrelativistic calculations \cite{12, 17,18,19,20, 21}, even though the theoretical bases are quite different.
\section{CONCLUSIONS}

A relativistic Dirac phenomenological calculation using an optical potential model could reproduce the experimental data for the excited states of the GSRB at the 1.047 GeV unpolarized proton inelastic scatterings from Ni isotopes, $^{62}$Ni and $^{64}$Ni reasonably well, achieving a little better agreement with the data compared to the results obtained in the norelativistic calculations.
 The Dirac equations are reduced to the Schr\"{o}dinger-like second-order differential equations to get the effective central and spin-orbit potentials, and surface-peaked phenomena are observed at the effective real central potentials for the scattering from $^{62}$Ni, as shown for the scatterings from $^{20}$Ne and $^{24}$Mg. The effective central potentials and the effective real spin-orbit potentials are found to have abnormal wiggling shape at about 3 fermi for the scattering from $^{62}$Ni, indicating some inner structure of the nucleus. The first-order rotational collective models are employed to accommodate the low-lying excited states of the GSRB in the nuclei, and the calculated deformation parameters are compared with those obtained for the other Ni isotopes.
 The multistep excitation via $2^+$ state is confirmed to be important for the $4^+$ state excitation of the GSRB at the proton scatterings from $^{62}$Ni and $^{64}$Ni, as previously shown at the proton scatterings from the other Ni isotopes, $^{58}$Ni and $^{60}$Ni. The Dirac phenomenological calculation for the inelastic scatterings from the Ni isotopes considering the first 3$^-$ excitation is also performed by using the first order vibrational collective model. It is found that the deformation parameters for the 2$^+$, the 4$^+$ states of the GSRB and the first 3$^-$ state agree pretty well with those obtained in the nonrelativistic calculations.

\begin{acknowledgments}
This work was supported by the research grant of the Kongju National University in 2018.
The author would like to thank Seong-Hyeon Jeong for his valuable technical help in the preparation of this paper.
\end{acknowledgments}


\begin{references}
%\begin{thebibliography}{}
\bibitem{1} L. G. Arnold, B. C. Clark, R. L. Mercer, and  P. Swandt, Phys. Rev. C {\bf 23}, 1949 (1981).
\bibitem{2} J. A. McNeil, J.  Shepard, and  S. J.  Wallace, Phys.  Rev. Lett  {\bf 50}, 1439 (1983); {\bf 50}, 1443 (1983).
\bibitem{3} S. Shim, Ph.D. dissertation, The Ohio State University 1989; L. Kurth, B. C. Clark, E. D. Cooper, S. Hama, S. Shim, R. L. Mercer, L. Ray, and G. W. Hoffmann, Phys.  Rev. C {\bf 49}, 2086 (1994).
\bibitem{4} S. Shim, B. C. Clark, E. D. Cooper, S. Hama, R. L. Mercer, L. Ray, J. Raynal, and H. S. Sherif, Phys. Rev. C {\bf 42}, 1592 (1990).
\bibitem{5} R. de Swiniarski, D. L. Pham, and J. Raynal, Z. Phys. A - Hadrons and Nuclei {\bf 343}, 179 (1992).
\bibitem{6} D. L. Pham and R. de Swiniarski, Nuovo Cimento A {\bf 107}, 1405 (1994).
\bibitem{7} J. J. Kelly, Phys. Rev. C {\bf71}, 064610 (2005).
\bibitem{8} S. Shim, M. W. Kim, B. C. Clark, and L. Kurth Kerr, Phys. Rev. C {\bf 59}, 317 (1999).
\bibitem{9} S. Shim, Shin-Ho Ryu and Min-Soo Kim, J. Korean. Phys. Soc. {\bf 51}, 271 (2007); S. Shim, Shin-Ho Ryu and Min-Soo Kim, J. Korean. Phys. Soc. {\bf 53}, 1146 (2008).
\bibitem{10} S. Shim and M. W. Kim, Int. Jou. of Mod. Phys. E {\bf 21}, 1250098 (2012).
\bibitem{11} S. Shim, to be published in Can. Jou. Phys. (2017).
\bibitem{12} P. Beuzit, J. Delaunay, J.P. Fouan and N. Cindro, Nucl. Phys. A {\bf 128}, 594 (1969).
\bibitem{13} J. Raynal, {\it Computing as a Language of Physics}, ICTP International Seminar Course, 281(IAEA, Italy, 1972); J. Raynal, {\it Notes on ECIS94}, Note CEA-N-2772, 1994.
\bibitem{14} S. Shim and M. W. Kim, J. Korean. Phys. Soc. {\bf 64}, 483 (2014).
\bibitem{15} S. Shim, Can. Jou. Phys. {\bf 95}, 317 (2017).
\bibitem{16} R. M. Lombard, G. D. Alkhazov and O. A. Domchenkov, Nucl. Phys. A {\bf 360}, 233 (1981).
\bibitem{17} l. Ray, T. Kozlowski, D.G. Madland, C.L. Morris, J.C. Pratt {\it et al}, Phys. Lett. {\bf 83B}, 275 (1979).
\bibitem{18} E. Fabrii, S. Micheletti, M. Pignanelli, F. G. Resmini, R. De Leo {\it et al}, Phys.  Rev. C  {\bf 21}, 844 (1980).
\bibitem{19} A. Ingemarsson, T. Johansson and G. Tibell, Nucl. Phys. A {\bf 365}, 426 (1981).
\bibitem{20} G.S. Kyle, N.M. Hintz, M.S. Oothoudt, M. Kaletka, P.M. Lang {\it et al}, Phys. Lett. {\bf 91B}, 353 (1980).
\bibitem{21} P.J. van Hall, S.D. Wassenaar, S.S. Klein, G.J. Nijgh, J.H. Polane and O.J. Poppema, J. Phys. G: Nucl. Part. Phys. {\bf 15}, 199 (1989).
%\end{thebibliography}
\end{references}
\end{document}